\documentclass[12pt,a4paper]{article}
\usepackage{amssymb,amsmath,latexsym}
\date{\today}

\title{Topological quantization of boundary forces and the integrated
  density of states}

\author{J. Kellendonk
\\
\\
{\small Institut Girard Desargues, 
Universit\'e Claude Bernard Lyon 1, F-69622 Villeurbanne}
\\
}

\date{\today}

\newcommand{\RR}{\mathbb R}

\newcommand{\CC}{\mathbb C}

\newcommand{\Uu}{{\cal U}}

\newcommand{\ZZ}{{\bf Z}}

\newcommand{\PP}{{\bf P}}

\newcommand{\TV}{{\tau}}
\newcommand{\TVh}{\hat{{\tau}}}

\newcommand{\hull}{\Omega}
\newcommand{\om}{\omega}

\newcommand{\CA}{$C^*$-algebra}

\newcommand{\cha}{\chi^{}}

\newcommand{\hsp}{\RR^{d-1}\times\RR^{\leq 0}}
\newcommand{\IDS}{\mbox{\rm IDS}}
\newcommand{\sv}[2]{\left(\begin{array}{c} #1 \\ #2 \end{array}\right)}

\begin{document}

\maketitle

\begin{abstract}
For quantum systems described by
Schr\"odinger operators on the 
half-space $\hsp$
the boundary force
per unit area and unit energy
is topologically quantised provided the Fermi energy lies in a gap of
the bulk spectrum. Under this condition it is also equal to the
integrated density of states at the Fermi energy. 
\end{abstract}

\section{Introduction}

Consider a quantum system on the half space $\hsp$.  
One distinguishes between its behaviour for $x_d<<0$ and $x_d$
near $0$ speaking about the bulk behaviour in the first, and
about the edge (or boundary) behaviour of the system in the second case. 
Bulk and edge behaviour are not independent but 
topologically quantised observables in the bulk
related to topologically quantised observables at the edge.
A famous example of this type is the quantum Hall
effect in which the Hall conductivity can either be related to a
current current correlation in the bulk or is simply the conductance
of the edge current 
\cite{Pru,Fro,SKR,KRS02,ElbauGraf02,KS2}
In this letter we present another example
relating the value of the integrated density of states on a gap (i.e.\
at energies lying in the gap) of the bulk spectrum to a force the
boundary exhibits on the edge states. 

We discuss in the next section the underlying model and provide a
proof of our claims in the case of one-dimensional periodic systems.
The proof for the general case will be given elsewhere.

\section{The model}

In line with recent descriptions of aperiodic systems 
\cite{Pastur80,Bel85,Bel92,KS1}
we describe the bulk behaviour of the system 
in the one particle approximation 
by a covariant family of Schr\"odinger
operators $\{H_\om\}_{\om\in\Omega}$,
$$H_\om=\frac{\hbar^2}{2m}\sum_{j=1}^d(\imath
\partial_j-\frac{q}{\hbar}A_j)^2 +V_\om\;,$$ 
all acting on $L^2(\RR^d)$,
where $A$ is a vector potential for an external
constant magnetic field and $\Omega$ is a space of disorder configurations
or of configurations which cannot be macroscopically distinguished.
The second case is interesting for ordered systems, like quasi-crystals,
in which $\hull$ can be the hull of a single point pattern describing
set of average positions of the atoms and possibly
decorated with information to distinguish
between the kind of atoms.
$\hull$ carries three structures. First, a metric
topology in which it is compact, namely two configurations are deemed
$\epsilon$-close if they agree up to an error
of order $\epsilon$ on a $\frac{1}{\epsilon}$-neighbourhood around
the origin $0\in\RR^d$ (see \cite{FHKphys} for a precise formulation). 
Second, an action of the 
group of translations which we denote by
$\om\mapsto x\cdot\om$ where $x\cdot\om$ is the translate of the 
configuration $\om\in\Omega$ which looks
around $x$ like $\omega$ around $0$. 
Then a family of potentials $\{V_\om\}_{\om\in\hull}$ is called
covariant if $V_\om(x-y)=V_{y\cdot\om}(x)$ which implies that 
the corresponding family of Schr\"odinger
 operators $\{H_\om\}_{\om\in\hull}$ is covariant w.r.t.\ magnetic
translations.    
Third, $\Omega$ comes with a translation invariant
Borel-probabiblity measure $\PP$
and all measurable quantities are averaged over $\Omega$ with respect
to this measure. This average has the meaning of a disorder average
but should also be carried out if the configurations in $\hull$
describe ordered systems, because they are supposed to be
macroscopically indistinguishable.
The union $\bigcup_{\om\in\hull}\sigma(H_\om)$ of all spectra $\sigma(H_\om)$
is called the bulk spectrum.

\subsection{Integrated density of states on gaps}

The integrated density of states at energy $E$
of a single operator $H_\om$ from the
family can be defined as the trace per unit volume $\TV$ of the spectral
projection $P_E(H_\om)$ of $H_\om$ onto the states up to energy $E$. 
If the probability measure $\PP$ is ergodic then 
the trace per unit volume of $P_E(H_\om)$ is
$\PP$-a.s.\ constant over $\hull$ and
\begin{equation}\label{eq-ids}
\IDS(E) = \int_\hull d\PP(\om) \TV(P_E(H_\om))\;.
\end{equation}
Thus the integrated density of states should be thought of as an
expectation value of the whole family of operators. 
The integrated density of states is constant on gaps in the bulk
spectrum. It has been shown that equation (\ref{eq-ids}) has a \CA ic
interpretation if $E$ lies outside the bulk spectrum.
In particular, the family $\{P_E(H_\om)\}_{\om\in\hull}$ 
can be viewed as a projection in the natural 
\CA\ associated with the configuration space
$\hull$ (and the magnetic field) and that the
l.h.s.\ of (\ref{eq-ids}) depends only on the homotopy class of $P_E$
in this \CA\ \cite{Bel92}. This formulation makes clear that
the value of the integrated density of states on a gap 
is topologically quantised, namely, first, it is
stable under perturbations of the Schr\"odinger operator by covariant
operators which do not lead to a closing the gap, and second, 
it lies in a countable
subgroup of $\RR$, the gap-labelling group, which depends only on the
topology of $\hull$, its measure $\PP$, and the translation action,
but not on the specific form of the potentials. 
In specific cases this group can be determined, e.g.\
for hulls of Delone sets of finite local complexity this group is the
sub-group generated by the relative frequencies of the patterns
appearing in the Delone set \cite{BBG,BO,KP}.

\subsection{Boundary force per unit area and unit energy in gaps}

To describe the behaviour near the edge we consider the same
family of operators but restrict them to 
$L^2(\RR^{d-1}\times\RR^{\leq  s})$ 
demanding Dirichlet boundary conditions at the boundary 
$\RR^{d-1}\times\{s\}$. 
We define the total boundary force to be minus the variation of
the energy under a variation of the position $s$ of the boundary. To
describe this in one and the same Hilbert space, say the one
corresponding to $s=0$, we 
vary instead the position of the configuration in the opposite direction.
Per unit area the boundary force exhibited on the states in
some interval $\Delta$ is therefore 
$$ \int_\hull d\PP(\om)\TVh (P_\Delta(\hat H_\om)\delta(\hat H_\om))\;,\qquad 
\delta (\hat H_\om) := \lim_{x_d\to 0} \frac{H_{x_de_d\cdot \om}}{x_d} =
\frac{\partial V_\om}{\partial x_d}\;,$$ 
where $\TVh$ is the trace per unit
area parallel to the boundary and
we have for clarity denoted the restriction of $H_\om$ to
$L^2(\RR^{d-1}\times\RR^{\leq  0})$  
with Dirichlet boundary conditions with a hat.
Note that we include the $\PP$-average directly into the definition
of the boundary force. This is crucial for its topological
quantisation
and it is not true that $\TVh (P_\Delta(\hat H_\om)\delta(\hat
H_\om))$ is $\PP$-a.s.\ constant over $\hull$ for ergodic measures.
The boundary force per unit area and energy at energy $E$ is then
\begin{equation}\label{eq-bforce}
F_b(E)=\lim_{\Delta\to\{E\}} \frac{1}{|\Delta|}\int_\hull d\PP(\om)
\TVh(P_\Delta(\hat H_\om)\frac{\partial V_\om}{\partial x_d})\;.
\end{equation}
$|\Delta|$ is the length of the interval which tends to $0$ in that limit.
The limit exists if $E$ lies in a gap of the bulk spectrum. Moreover,
if $E$ lies in a gap then $F_b(E)$ is a topological quantity. In fact,
it can be shown that 
\begin{equation}\label{eq-unitary} 
\frac{1}{|\Delta|}\int_\hull d\PP(\om)
\TVh(P_\Delta(\hat H_\om)\delta (\hat H_\om))
= 2\pi \TVh((\Uu_\om^*(\Delta)-1)\delta(\Uu_\om(\Delta)))
\end{equation}
where 
$$\Uu_\om(\Delta)-1 = 
\left(e^{2\pi\imath t (\hat H_\om-E_0)}-1\right) P_\Delta(\hat H_\om)
\;,\qquad t=  \frac{1}{|\Delta|}\;,\qquad E_0=\min\Delta\;.$$
Thus $\Uu_\om(\Delta)$ is essentially 
the time evolution of the states of energy
in $\Delta$ by the time which is the inverse of the width of $\Delta$.
Equation (\ref{eq-unitary}) can be interpreted in a \CA ic context and
shown to depend only on the homotopy class of the unitary
$\Uu_\om(\Delta)$. Therefore, $F_b(E)$ is topologically quantised in
the same way as the integrated density of states. In
fact, we can show using the tools of non-commutative topology of \CA s
developed in \cite{KS2} that, for energies $E$ in gaps and ergodic $\PP$,
 \begin{equation}
|F_b(E)|= \IDS(E).
\end{equation}
The proof of this result in the general case will be given in a
separate publication. 
In the next
section we give an elementary proof of the above equality for periodic
one-dimensional systems.

\section{One-dimensional periodic systems}

We consider in this section the probably simplest case, in which we
have a one-dimensional periodic configuration $\om_0$, of
period $L$, and $\Omega$ is the set
of its translates (no external magnetic field). Then 
$\Omega=\{x\cdot \omega_0|x\in\RR\}\cong \RR/L\ZZ$ with (normalised)
standard action of $\RR$ by translation and
Lebesgue measure. If we choose a differentiable periodic potential $V$ 
then $V_{\xi\cdot \om_0}(x)=V(x+\xi)$ 
defines a covariant family of potentials. 
For simplicity we write $V_\xi$ in place of $V_{\xi\cdot \om_0}$.
We need to combine results about the spectral theory of 
the family $H_\xi := -\partial^2 + V_\xi$, $\xi\in [0,L)$,
on three different spaces, see e.g.\ \cite{DS,Bel92} for
background information.
\begin{itemize}
\item[1)] On $L^2(\RR)$. Fixing $E\in\RR$ and $\xi\in [0,L)$ 
   one finds for each $a,b\in\CC$
   a unique solution of $(H_\xi-E)\Psi = 0$ with initial condition
   $\Psi(0)=a$, $\Psi'(0)=b$. $\Psi$ is a function over $\RR$
   which is not normalisable. Since $H_\xi$ is a linear operator these
   unique solutions define a linear map $\CC^2\to \CC^2$: 
   $\sv{\Psi(0)}{\Psi'(0)}\mapsto\sv{\Psi(L)}{\Psi'(L)}$ whose
   associated matrix
   is called the monodromy matrix, we denote it by $M(E,\xi)$. 
The spectrum of $H_\xi$ on $L^2(\RR)$ is the set
$\{E\in\RR| -2\leq  M_{11}(E,\xi)+M_{22}(E,\xi) \leq 2\}$. 
It is independent of $\xi$
   and hence equal to the bulk spectrum. It is convenient to call the
   closures of the connected components of the open set 
   $\{E\in\RR| -2< M_{11}(E,\xi)+M_{22}(E,\xi) < 2\}$ the bands. Then
   bands may touch.

\item[2)] On $L^2([0,L])$ with Dirichlet boundary conditions at the
   boundary points. On that space the spectrum of $H_\xi$ is a discrete
   countably infinite set $\{\mu_1(\xi),\mu_2(\xi),\cdots\}$, 
   ($\mu_j(\xi)<\mu_{j+1}(\xi)$). We call these spectral values
   Dirichlet eigenvalues of $H_\xi$. 
They are determined by the equation
$M_{12}(E,\xi)=0$. Between $\mu_j(\xi)$
   and $\mu_{j+1}(\xi)$ lies the $j+1$th band of the bulk spectrum
   (counted from the lowest band). 
   We call a Dirichlet
   eigenvalue $\mu_n(\xi)$ a left or right eigenvalue if its corresponding
   eigenfunction $\psi_{n,\xi}$ satisfies
   $|\psi'_{n,\xi}(L)|<|\psi'_{n,\xi}(0)|$ or
   $|\psi'_{n,\xi}(L)|>|\psi'_{n,\xi}(0)|$, respectively. The
   terminology comes from the exponential increase if the functions
   are considered over many periods
   which physically
   means that eigenfunctions of right eigenvalues are localised at the
   right edge. If two bands touch they touch at a Dirichlet eigenvalue
   which is neither a left nor a right eigenvalue.

   Sturm-Liouville theory gives us the important information that 
   a {\em real} eigenfunction $\psi_{n,\xi}$ of $H_\xi$ to $\mu_n(\xi)$ 
   has exactly $n$ zeroes on the half open
   interval $[0,L)$ (so the zero at $L$ is not counted). 

\item[3)] On $L^2(\RR^{\leq 0})$ with Dirichlet boundary condition at the
   boundary $0$. The spectrum of $H_\xi$ on that space is the union of the
   bulk spectrum with the right Dirichlet eigenvalues. In fact, $E$
   belongs to that spectrum iff the eigenvalues of $M(E,\xi)$ have
   modulus $1$ or $\sv{0}{1}$ is an eigenvector of
   $M(E,\xi)$ to an eigenvalue $\rho$ of modulus strictly
   larger than $1$. In the first case $E$
   belongs to a band and in the second the corresponding eigenfunction
   satisfies 
   $|\psi'_{n,\xi}(L)|=\rho|\psi'_{n,\xi}(0)|>|\psi'_{n,\xi}(0)|$.

   We give an alternative description of left or right Dirichlet eigenvalues,
   namely $\mu_n(\xi)$ is a left or right eigenvalue iff
   $\mu'_n(\xi)>0$ or $\mu'_n(\xi)<0$, respectively. In fact, we
   calculate $\mu'_n(\xi)$ for right eigenvalues 
\begin{equation}\label{eq-abl}
\mu'_n(\xi) = \partial_\xi
\int_{-\infty}^0 dx \overline{\hat\psi_{n,\xi}(x)} H_\xi \hat\psi_{n,\xi}(x)
= \int_{-\infty}^0 dx |\hat\psi_{n,\xi}(x)|^2 V'_\xi(x)
\end{equation}
where $\hat\psi_{n,\xi}$ is the {\em normalised} 
eigenfunction, $\int_{-\infty}^0
dx |\hat\psi_{n,\xi}(x)|^2 =1$. Clearly, this normalisation is only
possible since $\mu_n(\xi)$ is a right eigenvalue.
Using integration by parts and
$\hat\psi_{n,\xi}(0)=\hat\psi_{n,\xi}(-\infty)=\hat\psi_{n,\xi}'(-\infty)=0$
we find, 
\begin{eqnarray*}
 \int_{-\infty}^0 dx |\hat\psi_{n,\xi}(x)|^2 V'_\xi(x) & = &  
 -\int_{-\infty}^0 dx \overline{\hat\psi_{n,\xi}'(x)} 
\hat\psi_{n,\xi}(x) V_\xi(x) - c.c.\\
& = & - \int_{-\infty}^0 dx \overline{\hat\psi_{n_\xi}'(x)}
(E \hat \psi_{n,\xi}(x) +
 \frac{\hbar^2}{2m}\hat\psi_{n,\xi}''(x))-c.c. \\ 
& = &   -\frac{\hbar^2}{2m}  |\hat\psi_{n,\xi}'(0)|^2 < 0
\end{eqnarray*}
For left eigenvalues one proceeds similarly, but uses the space
$L^2(\RR^{\geq 0})$ instead. Since $0$ is then the left boundary of the
integral one obtains a relative minus sign in the calculation. The
remaining case is that $\mu_n(\xi)$ is neither a left nor a right
eigenvalue. Then it must be at a band edge and therefore an extremum
of $\mu_n$.

This has the following consequence which is crucial below:
Since a real eigenfunction to a 
right Dirichlet eigenvalue $\mu_n(\xi)$ has $n$ zeroes on
$[0,L)$ the equations $ \mu_n(\xi) =\mu$, $\mu_n'(\xi) < 0 $ have
exactly $n$ solutions on $\xi\in[0,L)$.
\end{itemize}

Using an approximation of the trace per unit volume by the trace per
unit volume in the representations $L^2([-NL,NL])$ with Dirichlet
boundary conditions one obtains from Sturm-Liouville theory 
in the limit $N\to\infty$
\begin{equation}\label{IDS} 
\IDS(\mu_n(\xi)) = \frac{n}{L}
\end{equation}  
independent of $\xi$.

We determine the boundary force (\ref{eq-bforce}).
If $\Delta$ is an interval in the $n$th gap of the 
bulk spectrum then $H_\xi$ has exactly one non-degenerate eigenvalue
(a right-Dirichlet eigenvalue) provided $\mu(\xi)\in\Delta$ and
$\mu'_n(\xi)<0$, otherwise it has none. 
Hence the integral kernel $\langle x|P_\Delta(\hat H_\xi)|y\rangle$
of $P_\Delta(\hat H_\xi)$ is 
$\hat\psi_{n,\xi}(x)\overline{\hat\psi_{n,\xi}(y)}
\cha_\Delta(\mu_n(\xi))\Theta(-\mu'_n(\xi))$ where $\cha_\Delta$ is
the characteristic function on the interval
$\Delta$ and $\Theta$ the Heavyside function.
In one dimension $\TVh$ is the operator trace and therefore
$$\TVh(P_\Delta(\hat H_\xi) V'_\xi) = 
\int_{-\infty}^0 dx
|\hat\psi_{n,\xi}(x)|^2\cha_\Delta(\mu_n(\xi))
\Theta(-\mu'_n(\xi)) V'_\xi(x)\;.$$
By (\ref{eq-abl}) and the fact that 
$ \mu_n(\xi) =\mu$, $\mu_n'(\xi) < 0 $ has
exactly $n$ solutions on $\xi\in[0,L)$
\begin{eqnarray*}
\int_0^L d\xi \int_{-\infty}^0 dx |\hat\psi_{n,\xi}(x)|^2
\cha_\Delta(\mu_n(\xi))\Theta(-\mu'_n(\xi)) V'_\xi & =& 
\int_0^L d\xi \mu_n'(\xi)\cha_\Delta(\mu_n(\xi))\Theta(-\mu'_n(\xi))\\
&=& -n \int_\Delta d\mu \quad = \quad - n |\Delta|.
\end{eqnarray*}
As it should be, this expression is negative and so the force points
into the sample. We conclude that 
$$ |F_b(E)|  = \frac{n}{L} $$
and so the strength of boundary force per unit area and unit
energy is equal to the integrated density of states for an energy
$E$ which lies in a gap of the bulk spectrum.
We end this section with some remarks.
\begin{enumerate}
\item
The integer $n$ appearing in the above expression may
also be interpreted as a winding number of the Dirichlet
eigenvalue on the complex spectral curve of $H_\xi$. This is similar
but not equal to the phenomenon observed in \cite{Hat93}.

\item
The above also yields, for arbitrary interval $\Delta$ in the
$n$th gap,
$$ \frac{\hbar^2}{2m |\Delta|}\int_0^L d\xi
|\hat\psi_{n,\xi}'(0)|^2 \cha_\Delta(\mu_n(\xi))\Theta(\pm\mu'_n(\xi)) =
\frac{\hbar^2}{2 m w_n}\int_0^L d\xi
|\hat\psi_{n,\xi}'(0)|^2\Theta(\pm\mu'_n(\xi))  = n\;. $$
Here $w_n$ is the width of the $n$th gap and, for the $+$-sign ($-$-sign),
$\hat\psi_{n,\xi}$
is a normalised eigenfunction to the left (right) Dirichlet eigenvalue. 
This equation seems interesting in its own right for
one-dimensional periodic operators. 

\item
Since, by the boundary conditions,  
$$|\hat\psi_{n,\xi}'(0)|^2=
\left. \frac{\partial_x^2|\hat\psi_{n,\xi}(x)|^2}{2}\right|_{x=0}$$
we see that the boundary force per unit area and unit
energy is determined by the $\PP$-average of the first non-vanishing
coefficient in the Taylor expansion of the density of the
particles at the edge.

\item
We note again that the $\PP$-average is crucial for the
topological quantisation. In fact, $\hat\psi_{n,\xi}'(0)$ tends to $0$
if $\xi$ tends to the extrema of the function $\mu_n(\xi)$.
\end{enumerate}


\begin{thebibliography}{99}

\bibitem[B85]{Bel85}
J. Bellissard, {\sl K-theory of C$^*$-algebras in solid
state physics},
in {\sl Statistical Mechanics and Field Theory: Mathematical Aspects},
{\sl Lecture Notes in Physics} {\bf 257}, edited by T. Dorlas, M. Hugenholtz,
M. Winnink, 99-156 (Springer-Verlag, Berlin, 1986).

\bibitem[B92]{Bel92}
J. Bellissard, {\sl Gap labelling theorems for Schr\"odinger
operators}, 538-630, in
{\sl From Number Theory to Physics},
(Springer, Berlin, 1992).

\bibitem[BBG]{BBG}
J. Bellissard, R. Benedetti, J. Gambaudo,
{\sl Spaces of tilings, finite telescopic approximations and
  gap-labelling},
preprint math.DS/0109062.

\bibitem[BO]{BO}
M. Benameur, H. Oyono-Oyono,
{\sl Gap-labelling for quasi-crystals (proving a conjecture by
  J. Bellissard)}, preprint math.KT/0112113.

\bibitem[DS88]{DS}
N. Dunford, J.T. Schwartz, {\sl Linear Operators, Part II: Spectral
  Theory}, 
(Wiley Classics Library Edition Published 1988, John Wiley and Sons,
New York, 1988). 

\bibitem[EG02]{ElbauGraf02}
P. Elbau, G.-M. Graf, {\sl Equality of bulk and edge Hall
conductance revisited}, 
Commun. Math. Phys. {\bf 229}, 415--432 (2002). 

\bibitem[FHK02]{FHKphys} A.H. Forrest, J.~Hunton, and J.~Kellendonk,
{\sl Cohomology of canonical projection tilings},
Commun. Math. Phys. {\bf 226}, 289--322 (2002).
                                                                   

\bibitem[F94]{Fro} J. Fr\"ohlich,
{\sl Mathematical aspects of the quantum Hall effect}, 
First European Congress of Mathematics, Vol. II (Paris, 1992), 
Progr. Math. {\bf 120}, 23-48
(Birkh\"auser, Basel, 1994). 

\bibitem[H93]{Hat93}
Y. Hatsugai,
{\sl Edge States in the integer quantum Hall effect and the Riemann
surface of the Bloch function},
Phys. Rev. B {\bf 48}, 11851-11862 (1993).
{\sl The Chern Number and Edge States in the Integer Quantum Hall
Effect},
Phys. Rev. Lett. {\bf 71}, 3697-3700 (1993).

\bibitem[KP]{KP}
J.~Kaminker, I.F.~Putnam, {\sl A proof of the Gap Labeling
  Conjecture}, preprint math.KT/0205102.

\bibitem[KRS02]{KRS02} J. Kellendonk, T. Richter, H. Schulz-Baldes, 
{\sl Edge current channels and Chern numbers in the integer quantum
Hall effect}, Rev. Math. Phys. {\bf 14}, 87-119 (2002).

\bibitem[KS03a]{KS1}
J. Kellendonk, H. Schulz-Baldes, {\sl Quantization of edge currents for
  continuous magnetic operators}, to appear in J.~Funct.~Anal..

\bibitem[KS03b]{KS2}
J. Kellendonk, H. Schulz-Baldes, {\sl Boundary maps for $C^*$-crossed
  products with $\RR$ with an application to the {Q}uantum {H}all {E}ffect},
submitted to Commun.~Math.~Phys..

\bibitem[Pr85]{Pru} A. M. M. Pruisken, {\sl Field Theory, Scaling and
the Localization Problem}, in
R. Prange, S. Girvin, Editors, {\sl The Quantum Hall Effect},
2nd Edition, (Springer-Verlag, Berlin, 1990).

\bibitem[Pa80]{Pastur80}
L. Pastur, {\sl Spectral properties of disordered systems in the
  one-body approximation}, Commun. Math. Phys. {\bf 75}, 179--196 (1980).

\bibitem[SKR00]{SKR} H. Schulz-Baldes, J. Kellendonk, T. Richter,
{\sl Simultaneous quantization of the edge and bulk Hall conductivity},
J. Phys. A: Math. Gen. {\bf 33}, L27-L32 (2000).

\end{thebibliography}
\end{document}